\documentclass{article}
\usepackage{spconf,amsmath,graphicx}

\usepackage{tikz}
\usepackage{pgfplots}
\pgfplotsset{compat=1.18} %
\newcommand{\vect}[1]{\ensuremath{\boldsymbol{\mathbf{#1}}}}
\usepackage{booktabs}
\usepackage{multirow}
\usepackage{siunitx}
\usepgfplotslibrary{groupplots}
\usepackage{pgfmath}
\usepackage{amsfonts}
\usepackage{setspace}

\usepackage{stackengine}
\usetikzlibrary{shapes.geometric, calc, fit, positioning, patterns}
\usepackage[acronym]{glossaries}
\newacronym{ode}{ODE}{Ordinary Differential Equation}
\newacronym{mfa}{MFA}{Montreal Forced Aligner}
\newacronym{tts}{TTS}{Text-to-Speech}
\newacronym{vae}{VAE}{Variational Autoencoder}
\newacronym{elbo}{ELBO}{Evidence Lower Bound}
\newacronym{cnf}{CNF}{Continuous Normalizing Flow}
\newacronym{eer}{EER}{Equal Error Rate}
\newacronym{wer}{WER}{Word Error Rate}
\newacronym{fp}{FP}{False Positives}
\newacronym{fn}{FN}{False Negatives}
\newacronym{asr}{ASR}{Automatic Speech Recognition}
\newacronym{pvqd}{PVQD}{Perceptual Voice Qualities Database}
\newacronym{ve}{VE}{Voice Editing}
\newacronym[longplural=Perceptual Voice Qualities]{pvq}{PVQ}{Perceptual Voice Quality}
\newacronym{cape-v}{CAPE-V}{Consensus Auditory-Perceptual Evaluation of Voice}
\newacronym{mos}{MOS}{Mean Opinion Score}
\newacronym{smos}{SMOS}{Speaker Similarity MOS}
\newacronym{ccnf}{CCNF}{Conditional Continuous Normalizing Flow}
\newacronym{hnr}{HNR}{Harmonic-to-Noise Ratio}
\newacronym{cpp}{CPP}{Cepstral Peak Prominence}
\newacronym{vad}{VAD}{Voice Activity Detection}
\newacronym{asv}{ASV}{Automatic Speaker Verification}

\usepackage[colorlinks=false]{hyperref}
\usepackage{cleveref}
\newcommand{\loga}{\mathrm{log}}
\title{Disentangling Pitch and Creak for Speaker Identity Preservation in Speech Synthesis}
\name{Frederik Rautenberg$^1$, Jana Wiechmann$^2$, Petra Wagner$^2$, Reinhold Haeb-Umbach$^1$} %
\address{$^1$Paderborn University, Paderborn, Germany; $^2$Bielefeld University, Bielefeld, Germany}

\newcommand{\meanone}[1]{%
  \ifdim #1 pt < 0 pt
    \pgfmathparse{abs(#1)}
    -\num[round-mode=places, round-precision=1]{\pgfmathresult}
  \else
    \phantom{-}\num[round-mode=places, round-precision=1]{#1}
  \fi
}

\newcommand{\meantwo}[1]{%
  \ifdim #1 pt < 0 pt
    \pgfmathparse{abs(#1)}
    -\num[round-mode=places, round-precision=2]{\pgfmathresult}
  \else
    \phantom{-}\num[round-mode=places, round-precision=2]{#1}
  \fi
}

\newcommand{\meanexp}[1]{%
    \num[round-mode=figures,
         round-precision=1,
         round-integer-to-decimal=false,
         scientific-notation=true,
         output-exponent-marker = e]{#1}%
}

\begin{document}
\maketitle
\begin{abstract}
We introduce a system capable of faithfully modifying the perceptual voice quality of \textit{creak} while preserving the speaker's perceived identity. While it is well known that high \textit{creak} probability is typically correlated with low pitch, it is important to note that this is a property observed on a population of speakers but does not necessarily hold across all situations. Disentanglement of pitch from \textit{creak} is achieved by augmentation of the training dataset of a speech synthesis system with a speaker manipulation block based on conditional continuous normalizing flow. The experiments show greatly improved speaker verification performance over a range of \textit{creak} manipulation strengths. 
\end{abstract}
\begin{keywords}
Voice Modification, Speech Synthesis
\end{keywords}
\section{Introduction}
In recent years, interest in voice editing has grown, including the manipulation of emotions \cite{triantafyllopoulos2023overview}, as well as the manipulation of so-called \gls{pvq} \cite{netzorg2023permod, lameris25_interspeech} features for a better understanding of voice characteristics. In this study, we focus on manipulating the voice quality \textit{creak} \cite{lameris24_interspeech, rautenberg25_interspeech}. \textit{Creaky voice} is typically characterized by a low rate of vocal fold vibration combined with a constricted glottis, resulting in a lower and irregular pitch \cite{keating2015acoustic}. 
As shown in \cite{rautenberg25_interspeech}, a global manipulation method can be used to modify non-persistent voice qualities, including \textit{creak}. Experiments with speech experts demonstrated that the perceived strength of that quality in the synthesized voice matched with the intended conditioning. The approach in \cite{rautenberg25_interspeech} employs the \gls{ccnf} concept \cite{grathwohl2018ffjord} to learn a \textit{creak}-conditioned data distribution of speaker representations. 

Because the data distribution is learned, the model captures all correlations present in the training data. 
In \cite{rautenberg25_interspeech}, the LibriTTS-R \cite{koizumi2023libritts} dataset was pseudo-labeled using CreaPy \cite{paierl2023creapy}. Since the dataset does not contain speakers actively altering their voice quality, each utterance from a speaker is consistently labeled with a similar \textit{creak} probability. This causes the \textit{creak} probability to be negatively correlated with the speaker's fundamental frequency, the pitch. We call this correlation ``inter-speaker'' creak-to-pitch correlation, because it indicates that speakers with high \textit{creak} probability tend to have lower pitch. This is not surprising, as, according to \cite{keating2015acoustic}, the prototypical form of \textit{creak} has exactly this property.  

However, there is a range of applications, where we are interested in modifying the degree of a PVQ for a given speaker, i.e., modifying the \textit{creak} probability, while preserving the speaker identity. For this, the above ``inter-speaker'' correlation observed in a population of speakers is unfavorable, because a speaker with a modified \textit{creak} probability would be identified to be a different speaker due to the also modified pitch. To train a system capable of studying ``intra-speaker'' variability, a dataset is required in which speakers vary their voice along a \gls{pvq} axis. Such datasets, however, are almost non-existent. An exception is the GTR-Voice dataset
published in \cite{li2024articulatory}. It consists of recordings of a voice actor who manipulates his voice along three articulatory-phonetic dimensions: \textit{Glottalization}, \textit{tenseness}, and \textit{resonance}. This enables the study of ``intra-speaker'' correlations of acoustic features along each dimension. However, such an approach requires skilled voice actors capable of systematically varying their voice along these axes, which is expensive, and some voice qualities are difficult to trigger. 
There exists no dataset in which speakers explicitly manipulate their voice along the \textit{creak} dimension. 

In this contribution, we propose an alternative, which allows the manipulation of \textit{creak} probability while maintaining the perceived speaker identity. We apply pitch shifting to the LibriTTS-R dataset to break the ``inter-speaker'' creak-to-pitch correlation. We show that training with the augmented dataset enables the model to disentangle pitch from \textit{creak}, resulting in more faithful ``intra-speaker'' \textit{creak} manipulation, which improves speaker preservation after voice manipulation\footnote{\href{https://groups.uni-paderborn.de/nt/icassp_2026_creak_manipulation/norm_flow.html}{go.upb.de/pitch-creak-disentanglement}}, while other acoustic correlates remain appropriately aligned with \textit{creak} conditioning.

This work aims to support speech experts in teaching voice quality perception \cite{wiechmann2025challenges}. A key requirement is to manipulate a target voice quality in synthesized speech without altering the speaker identity. Beyond that, such controlled manipulation enables data augmentation: it can improve \gls{asr} robustness to disordered speech \cite{zheng2025interspeech} and mitigate \gls{asv} degradation for dysphonic speakers \cite{tayebi2023effect}.

\section{System Description}
\begin{figure}[!b]
    \centering
    \vspace*{-0.5cm}
    \tikzstyle{point} = [draw, fill=white, circle, radius=1cm, scale=0.5]
\tikzstyle{smallpoint} = [draw, fill=black, circle, radius=1cm, scale=0.3]
\tikzstyle{block} = [rectangle, draw, text centered, minimum width=1.7cm, minimum height=0.6cm, rounded corners, fill=orange!30]
\tikzstyle{smallblock} = [rectangle, draw, text centered, minimum width=0.3cm, minimum height=0.5cm, rounded corners, inner sep=2pt, fill=orange!30]
\tikzstyle{encoder} = [trapezium, draw,  text centered, minimum width=1.7cm, minimum height=0.6cm, fill=blue!20]

\def\groupOneYOffset{0.9}
\def\groupTwoYOffset{1.2}  
\def\groupThreeYOffset{2.1} 
\def\smallBlockSpacing{0.4}

\begin{tikzpicture}[auto, node distance=0.5cm, remember picture]
    \node [] (input_text) {$\vect{c}_{\mathrm{text}}$};
    \node[encoder, above = of input_text] (TE) {\stackanchor{Text}{Encoder}};
    \node[block, above = 1cm of TE] (AG) {Projection};

    \node[block, above = 0.5cm of AG] (frame) {\stackanchor{Feature}{Extraction}};

    \node[block, above = 0.5cm of frame] (FB) {Flow};
    \node[block, right = 0.5cm of AG] (DP) {\stackanchor{Duration}{Predictor}};
    \draw[->, -stealth, thick, rounded corners] (DP.north) |- ([shift=({0.05,0})] frame.east);
    \draw[->, -stealth, thick] (AG.north) --([shift=({0,-0.05})]frame.south);
    \node[encoder, shape border rotate = 180, fill=green!30] at ([shift=({0.3,1})] FB.north |- FB.north) (hifi) {\stackanchor{HiFi-GAN}{Generator}};
    \node[above=of hifi] (speech_signal_output) {$\vect{\tilde{x}}$};

    \draw[->, -stealth, thick] (input_text) -- ([shift=({0,-0.05})]TE.south);
    \draw[->, -stealth, thick] (TE) -- ([shift=({0,-0.05})]AG.south);
    \draw[->, -stealth, thick] (FB) -- node[left] {$\vect{Z}$} ([shift=({-0.3,-0.05})]hifi.south);
    \draw[->, -stealth, thick] (hifi.north) -- ([shift=({0,-0.05})]speech_signal_output.south);
    \draw[->, -stealth, thick] (frame.north) -- ([shift=({0,-0.05})]FB.south);

    \node[encoder, right = 2cm of TE] (SE) {\stackanchor{Speaker}{Encoder}};
    \node[block, rounded corners] at (SE.north |- frame.west) (ode) {\stackanchor{Manipulation}{Block}};
    \node at (input_text.east -| SE.south) (speech_signal){$\vect{x}$};

    \draw[->, -stealth, thick] (speech_signal) -- ([shift=({0,-0.05})] SE.south);

    \node[smallpoint, above=0.4cm of TE] (sm2) {};
    \node[smallpoint] at (DP.east -| SE.north) (sm1) {};
    \draw[->, -stealth, thick, rounded corners] (sm1) -- ([shift=({0.05, 0})] DP.east);
    \draw[->, -stealth, thick] (SE.north) -- node[right, near start] {$\vect{s}$} ([shift=({0,-0.05})] ode.south);
    \draw[->, -stealth, thick, rounded corners] (sm2) -| ([shift=({0,-0.05})]DP.south);
    \node at ([shift=({0.8,0.2})]ode.east) (attribut){$\vect{\tilde{a}}$};
    \draw[->, -stealth, thick] (attribut.west) -- ([shift=({0.05, 0.2})]ode.east);
    \node at ([shift=({0.8,-0.2})]ode.east) (attribut){$\vect{a}$};
    \draw[->, -stealth, thick] (attribut.west) -- ([shift=({0.05, -0.2})]ode.east);

    \draw[->, -stealth, thick, rounded corners] (ode.north) |-  node[right, near start] {$\vect{\tilde{s}}$} (FB.east);

    \node[smallpoint] at (FB.east -| DP.north) (sm2) {};
    \draw[->, -stealth, thick, rounded corners] (sm2.north) |- ([shift=({0.4,0.15})] hifi.south |- FB.north) -- ([shift=({0.4,-0.05})]hifi.south);
\end{tikzpicture}
    \vspace*{-0.3cm}
    \caption{TTS inference with a speaker embedding manipulation block, where $\vect{a}$ is the creak probability of $\vect{x}$ and $\vect{a} + \vect{\tilde{a}}$ its modified strength}
    \label{fig:tts}
\end{figure}
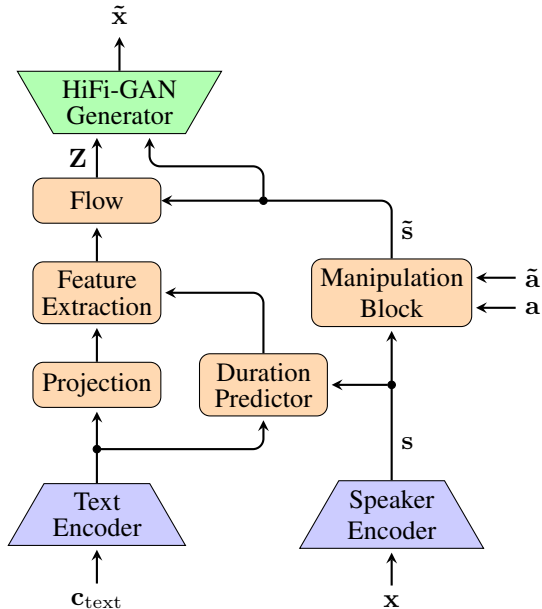
In \cite{anastassiou2024voiceshop}, a global speaker manipulation module, i.e., a module that alters the entire speaker embedding at once rather than modifying local or frame-level characteristics, was integrated into a voice conversion system to edit a speaker’s voice while preserving the identity. This manipulation was realized through a \gls{ccnf} \cite{abdal2021styleflow}. Building on this idea, \cite{rautenberg2025speech} introduced \gls{pvq} manipulation by incorporating a \gls{ccnf}-based manipulation block into a \gls{tts} system. The underlying \gls{tts} model is YourTTS \cite{casanova2022yourtts, kim2021conditional}. More recently, \cite{rautenberg25_interspeech} extended the approach to voice quality manipulation, focusing on the non-persistent attribute \emph{creak}.  

In this work, we follow the approach of \cite{rautenberg25_interspeech}. Our overall inference system is illustrated in \Cref{fig:tts}. Speech synthesis is performed with YourTTS \cite{casanova2022yourtts}, where training is based on maximizing the \gls{elbo} \cite{kim2021conditional}:
\begin{equation}
    \begin{aligned}
        \loga \, p_{\vect{X}}(\vect{x} | \vect{c}) \geq  \mathbb{E}_{q_{\vect{Z}}} \left[ \vphantom{\frac{test}{test}} \right.& \loga \,p_{\vect{X}}(\vect{x} | \vect{z}) - \left. \loga\frac{q_{\vect{Z}}(\vect{z}|\vect{x})}{p_{\vect{Z}}(\mathbf{z}|\vect{c})} \right] \, ,  
    \end{aligned}
\end{equation}
where $\vect{z} \in \mathbb{R}^{D \times T}$ denotes the latent embedding, $\vect{x}$ the speech signal, and $\vect{c} = [ \vect{c}_{\mathrm{text}}, \vect{s}]$ the conditioning, which consists of the text embedding $\vect{c}_{\mathrm{text}}$ and the speaker embedding $\vect{s}$. The prior distribution is given by $p_{\vect{Z}}(\vect{z}|\vect{c})$, the likelihood by $p_{\vect{X}}(\vect{x}|\vect{z})$, and the posterior by $q_{\vect{Z}}(\vect{z}|\vect{x})$. All distributions are modeled with neural networks. The prior encoder is only required during training.  

To perform a global manipulation of speaker attributes, we employ a manipulation block based on a \gls{ccnf}, which operates on the speaker embedding $\vect{s}$. The role of the normalizing flow is to model the conditional distribution of speaker embeddings given their attribute vectors. The conditioning vector $\vect{a}$ encodes the strength levels of the \glspl{pvq}. Using the normalizing flow, new speaker embeddings $\vect{\tilde{s}}$ can then be generated for a desired conditioning $\vect{\tilde{a}}$.

The normalizing flow is trained with the training dataset $\mathcal{S} = \left\{\vect{s}_n\right\}_{n=1}^N$ by maximizing the following log-likelihood function\cite{grathwohl2018ffjord}:
\begin{align}
    &l = \sum_n \loga p_{\vect{S}} \left( \vect{s}_n | \vect{a}_n \right) \nonumber \\
    &= \sum_n \loga p_{\vect{Z}_0}\left( \vect{z}_n(t_0)\right) + \int_{t_1}^{t_0}\mathrm{tr}\left( \frac{\mathrm{d} g(\vect{z}(t), t, \vect{a}_n)}{\mathrm{d}\vect{z}(t)} \right) \mathrm{d}t
    \label{eq:likelihood_2}
\end{align}
with 
\begin{align}
    \label{eq:transformation}
    \vect{z}_n(t_0) = \vect{z}_n(t_1) + \int_{t_1}^{t_0}  g(\vect{z}(t), t, \vect{a}_n) \mathrm{d}t  \, .
\end{align}

Here, $\vect{z}_n(t_0) \sim \mathcal{N}(\mathbf{0}, \mathbf{I})$, $g(\cdot)$ denotes a parameterized function, and the initial condition is given by $\vect{s}_n = \vect{z}_n(t_1)$. Computing the log-likelihood requires solving two \gls{ode} problems, namely \Cref{eq:likelihood_2} and \Cref{eq:transformation}. New speakers with modified attributes can be obtained through the following procedure. First, the embedding $\vect{s}_n$ and its attribute vector $\vect{a}_n$ are extracted from a speech signal $\vect{x}_n$. The embedding is then mapped to a latent representation $\vect{z}_n(t_0)$ by solving the \gls{ode} in \Cref{eq:transformation}, initialized with $\vect{s}_n = \vect{z}_n(t_1)$. To introduce the speaker changes, the inverse transformation is solved with the initial condition $\vect{z}_n(t_0)$, but with the attribute vector shifted to $\vect{\tilde{a}}_n$. This produces the manipulated embedding $\tilde{\vect{s}}_n$ which will serve as input for the \gls{tts} system.

\section{Dataset adaptation}
\begin{table}[!b]
\vspace{-0.5cm}
  \caption{Correlation coefficients $R$ and slopes $\alpha$ between pitch and creak probability on LibriTTS-R before (Set 1) and after adaptation (Set 2)}
  \vspace{0.25cm}
  \label{tab:correlation_coef}
    \setlength{\tabcolsep}{4pt} %
  \centering
  \begin{tabular}{c c c c c c c c}
    \toprule
    Set & \multicolumn{2}{c}{Overall} & \multicolumn{2}{c}{Male} & \multicolumn{2}{c}{Female} \\
    & $R$ & $\alpha$ & $R$ & $\alpha$ & $R$ & $\alpha$\\ 
    \midrule
    1  & \meanone{-0.75} & \meanexp{-0.0022} & \meanone{-0.67} & \meanexp{-0.004} & \meanone{-0.38} & \meanexp{-0.001} \\
    2 & \meanone{-0.58} & \meanexp{-0.002} & \meanone{-0.03} & - & \meanone{0.04} & -  \\
    \bottomrule
  \end{tabular}
\end{table}

\begin{figure}[!t]
    \newcommand{\plotwidth}{8.cm}
    \newcommand{\plotheight}{3.8cm}
    \centering
    \input{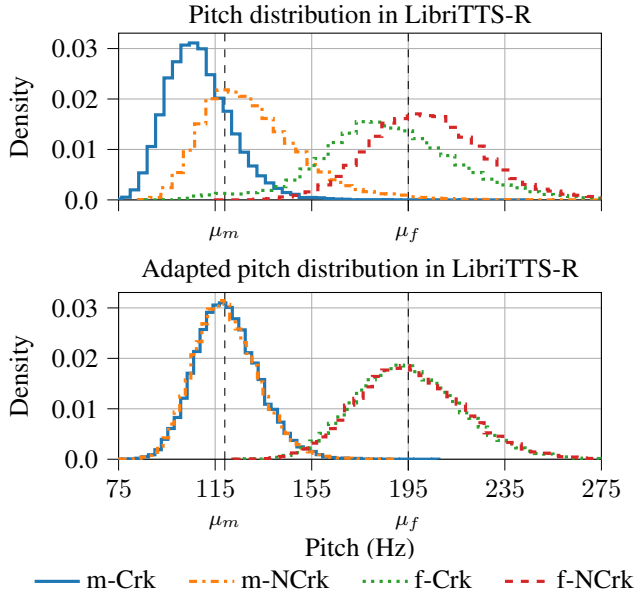}
    \vspace{-0.7cm}
    \caption{Distribution of pitch for male (m) and female (f) speakers, where samples are grouped into \textit{creak} (Crk) and non-\textit{creak} (NCrk) categories, before and after adaptation}
    \label{fig:distribution_pitch}
    \vspace{-0.35cm}
\end{figure}

For training, we used the LibriTTS-R \cite{koizumi2023libritts} dataset, with around 585 hours of speech data with 2,456 speakers. First, we applied a \gls{vad} on the audio samples, then we used CreaPy \cite{paierl2023creapy} to label the dataset with \textit{creak} probabilities and \cite{morrison2023cross} to estimate the mean pitch of each utterance. 

\Cref{fig:distribution_pitch} shows the distribution of mean pitch, categorized as \textit{creak} or non-\textit{creak}. Thresholds per gender were set from the dataset means: 0.5 (male), 0.3 (female). Across both genders, higher pitch values are associated with lower \textit{creak} probabilities, and vice versa. \Cref{tab:correlation_coef} reports the Pearson correlation coefficient $R$, confirming a negative correlation, particularly for male speakers. This observation is consistent with the prototypical description of \textit{creaky} voice reported in \cite{keating2015acoustic}. 

The fact that both features are correlated, doesn't tell by how much the pitch changes if the \textit{creak} probability increases. The slope $\alpha$ of the linear correlation between \textit{creak} and pitch, which describes this, has been determined on the training data set by using a linear interpolation. Note that this is the slope of the ``inter-speaker'' correlation, indicating the relationship of a speaker's \textit{creak} and pitch, measured on a set of speakers. However, for our application we are interested in modifying the \textit{creak} property of a given speaker's voice, while at the same time preserving its perceived identity. Note, that no datasets exist that contain the same speaker's voice at different levels of \textit{creak}. Thus, this ``intra-speaker'' creak-to-pitch correlation is unknown. 

We found, however, in informal experiments that the slope of the creak-pitch correlation in the training data set negatively affected the speaker identity preservation and thus needed to be reduced significantly. This asks for breaking the correlation between the two, effectively disentangling pitch from \textit{creak}, in order to keep speaker identity at different levels of \textit{creak}. The slopes reported in \Cref{tab:correlation_coef} indicate that increasing \textit{creak} probability has a measurable effect on pitch. For example, a full creaky male speaker would experience a pitch increase of approximately $250\,\mathrm{Hz}$ across the full range of \textit{creak} probability.

Our approach is to break this correlation by a simple dataset adaptation. First, we calculated the mean pitch for both genders (male: $119\,\mathrm{Hz}$ and female $195\,\mathrm{Hz}$). Then, the mean pitch of an utterance was shifted to the gender's mean value to remove the pitch correlation. We calculated the pitch difference $\Delta$ in semitones 
\begin{equation}
    \Delta = 12 \cdot \mathrm{log}_2\left( \frac{\Bar{f}_c}{\Bar{f}_{n,c}} \right) \, ,
\end{equation}
where $\Bar{f}_c$ is the mean pitch of the gender class and $\Bar{f}_{n,c}$ the mean pitch of the utterance $\vect{x}_n$. Then we calculated the new pitch contour as
\begin{equation}
    \tilde{f}_{n,c} = f_{n,c} \cdot 2^{\frac{\Delta + \mathrm{u} \cdot b} {12}} \, 
\end{equation}
where $u\sim\mathcal{N}(0,1)$ and $b$ is the target standard deviation, $f_{n,c}$ is the old pitch contour. The pitch of the utterance $\vect{x}_n$ is now shifted to the new contour $\tilde{f}_{n,c}$ by using TD-PSOLA \cite{moulines1990pitch}\footnote{https://github.com/maxrmorrison/psola}. We added noise to further reduce the correlation in the training data and introduce randomness to the results, in our experiments we set $b=2$. After applying the manipulation, the updated \textit{creak} probabilities were extracted using CreaPy. 

\Cref{fig:distribution_pitch} shows the distribution of the adapted utterances $\vect{\hat{x}}_n$, where it is evident that within each gender class, the distributions are now largely independent of pitch. This effect is also reflected in the correlation coefficients reported in \Cref{tab:correlation_coef}. While the within-gender pitch-to-creak correlation is almost entirely removed, an overall correlation remains, likely due to female speakers generally exhibiting less creaky voice than male speakers.

\section{Voice quality control}
We trained the \gls{tts} system on the LibriTTS-R \cite{koizumi2023libritts} dataset until convergence, using a batch size of 50 and a learning rate of $1\text{e-}4$. The \textit{creak} manipulation block, implemented as a \gls{ccnf}, was trained separately to maximize the log-likelihood objective in \Cref{eq:likelihood_2}, using a batch size of 200 and the same learning rate. The normalizing flow function $g(\cdot)$ was realized with a single CCNF block \cite{abdal2021styleflow} with a hidden dimensionality of 512. 

We trained three normalizing flows: the \textit{base}-flow \cite{rautenberg25_interspeech} on LibriTTS-R, the \textit{adapted}-flow on the pitch augmented LibriTTS-R dataset, and a combined flow trained on both datasets, which we refer to as \textit{combined}-flow. The conditioning vector $\vect{a}_n$ encodes the estimated strength of six \glspl{pvq} for a given input utterance $\vect{x}_n$. \textit{Breathiness}, \textit{roughness}, \textit{resonance}, and \textit{weight} were estimated with the random forest regressors from \cite{rautenberg2025speech}, mean pitch was extracted using \cite{morrison2023cross}, and \textit{creak} probability was obtained with CreaPy \cite{paierl2023creapy}. 
Prior to feature extraction, we applied \gls{vad} to remove non-speech segments. In informal experiments, we found that adding pitch as conditioning alone was insufficient for disentanglement.
To create manipulated conditioning vectors $\vect{\tilde{a}}_{n, \beta}$, we started from the original vectors $\vect{a}_n$ and shifted only the \textit{creak} dimension with factors $\beta \in [-1.25, 1.25]$ in steps of $0.25$. The resulting vectors were then used to synthesize the manipulated utterances $\vect{\tilde{x}}_{n,\beta}$. As shown in \cite{rautenberg25_interspeech}, this approach produced changes in perceived creakiness that matched well with expert listener judgments, while preserving naturalness.  
Here, our focus lies on two aspects: (i) whether acoustic correlates of \textit{creak} modification remain consistent after removing pitch dependencies, and (ii) whether speaker identity is preserved during manipulation.

\subsection{Acoustic Feature Analysis}

\begin{table}[!b]
  \centering
  \vspace{-0.65cm}
  \caption{Slope $\alpha$ and correlation coefficient $R$ for each acoustic metric. $\downarrow$ indicates the expected negative and $\uparrow$ the expected positive correlation\cite{paierl2023creapy, keating2015acoustic, davidsonCREAK2021, heldner2019voice} between the acoustic feature and increasing creaky voice.}
  \vspace{0.25cm}
  \setlength{\tabcolsep}{4pt}
  \begin{tabular}{l@{\hskip 1pt} c c | c c | c c}
    & \multicolumn{2}{c}{\textit{base}-flow} & \multicolumn{2}{c}{\textit{adapted}-flow} & \multicolumn{2}{c}{\textit{combined}-flow} \\
    \cmidrule(l){2-3} \cmidrule(l){4-5} \cmidrule(l){6-7}
    Metric & $\alpha$ & $R$ & $\alpha$ & $R$ & $\alpha$ & $R$ \\
    \midrule
    $\Delta$ pitch $\downarrow$   & \meantwo{-47.03675}   & \meanone{-0.990} & \meantwo{-2.62871}    & \meanone{-0.682} & \meantwo{-11.24786} & \meanone{-0.990} \\
    $\Delta$ h1h2 $\downarrow$    & \meanexp{-0.00035} & \meanone{-0.976} & \meanexp{-0.00038} & \meanone{-0.957} & \meanexp{-0.00040} & \meanone{-0.965} \\
    $\Delta$ hnr $\downarrow$     & \meantwo{-3.97169}    & \meanone{-0.988} & \meantwo{-1.68805} & \meanone{-0.979} &\meantwo{-2.30426} & \meanone{-0.988} \\
    $\Delta$ cpp $\downarrow$     & \meantwo{-0.37263}    & \meanone{-0.866} & \meantwo{-0.77972}    & \meanone{-0.966} & \meantwo{-0.67543} & \meanone{-0.920} \\
    $\Delta$ creak $\uparrow$     & \meantwo{0.25494}   & \meanone{0.993}  & \meantwo{0.14372}   & \meanone{0.996}  & \meantwo{0.18774} & \meanone{0.993} \\
  \end{tabular}
  \label{tab:acoustic_corr}
\end{table}

For the acoustic investigation, we analyzed the same features used by CreaPy \cite{paierl2023creapy} for \textit{creak} detection: mean pitch, H1-H2 (the strength of the first harmonic/fundamental relative to the second harmonic) and \gls{hnr}. 
Additionally, we investigated the \gls{cpp}, which is the amplitude of the first harmonic relative to the regression line in the power cepstrum. CPP characterizes the signal harmonicity. In line with existing empirical research \cite{keating2015acoustic, wiechmann2025challenges, davidsonCREAK2021, heldner2019voice}, we expect creaky voice to lead to 
 lower pitch, lower H1-H2, lower \gls{hnr} and lower \gls{cpp}. %

All features were extracted from synthesized speech $\vect{\tilde{x}}_{n,\beta}$ after applying \gls{vad}, using Parselmouth \cite{jadoul2018introducing}, a Python interface to Praat\footnote{www.praat.org}. 
To obtain the correlation slopes $\alpha$, we generated 4,000 random utterances from the test set across different $\beta$ values and applied linear interpolation between acoustic features and \textit{creak} conditioning.

As shown in \Cref{tab:acoustic_corr}, all models exhibit a negative correlation between pitch and \textit{creak} conditioning, but \textit{adapted}-flow and \textit{combined}-flow show considerably smaller slopes compared to the baseline. The expected correlates of creaky voice (H1-H2, HNR, CPP) remained consistently linked to \textit{creak}, though with slightly lower slopes in some cases. This suggests that the manipulation is now more robust against pitch artifacts, while minor reductions may partly stem from the TD-PSOLA based adaptation. The \textit{creak} prediction still reflects the manipulations, but the overall change in probability is slightly smaller. Note that mean pitch is also one of the features used by CreaPy for \textit{creak} prediction. \textit{Combined}-flow yields slightly higher slopes than the \textit{adapted}-flow, indicating a stronger change in \textit{creak} probability.

\subsection{Speaker verification under creak manipulation}

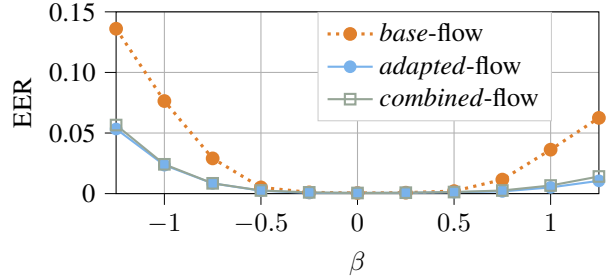
\begin{figure}
    \centering
    \newcommand{\plotwidth}{8cm}
    \newcommand{\plotheight}{4.cm}
    \begin{tikzpicture}

\definecolor{darkgray176}{RGB}{176,176,176}
\definecolor{lightgray204}{RGB}{204,204,204}

\definecolor{peru22412844}{RGB}{224,128,44}
\definecolor{skyblue121178235}{RGB}{121,178,235}
\definecolor{sage150165150}{RGB}{150,165,150} %

\begin{axis}[
legend cell align={left},
legend style={fill opacity=0.8, draw opacity=1, text opacity=1, draw=lightgray204},
tick align=outside,
tick pos=left,
x grid style={darkgray176},
xlabel={$\beta$},
xmajorgrids,
xmin=-1.25, xmax=1.25,
xtick style={color=black},
y grid style={darkgray176},
ylabel={EER},
ymajorgrids,
ymin=0, ymax=0.15,
ytick style={color=black},
ytick={0, 0.05, 0.1, 0.15},
yticklabels={0, 0.05, 0.10, 0.15},
width=\plotwidth, height=\plotheight,
legend style={
  fill opacity=0.8,
  draw opacity=1,
  text opacity=1,
  at={(0.66,1.)}, %
  anchor=north,    
},
]
\addplot [very thick, peru22412844, mark=*, mark size=2, mark options={solid}, dotted]
table {%
-1.25 0.136059670781101
-1 0.0763888888888888
-0.75 0.0290637860080684
-0.5 0.00514403292224601
-0.25 0.00102880658426801
0 0.000514403292580253
0.25 0.000771604938266154
0.5 0.00205761316872343
0.75 0.0115740740743383
1 0.0362654320983685
1.25 0.0624999999997258
};
\addlegendentry{\textit{base}-flow}
\addplot [thick, skyblue121178235, mark=*, mark size=2, mark options={solid}]
table {%
-1.25 0.0533453887884268
-1 0.0235081374321869
-0.75 0.00843881856540892
-0.5 0.00241109101872021
-0.25 0.000602772755034091
0 0.000602772754671411
0.25 0.000602772755616766
0.5 0.00120554550991263
0.75 0.00180831826415064
1 0.00512356841527025
1.25 0.0105485232070514
};
\addlegendentry{\textit{adapted}-flow}

\addplot [thick, sage150165150, mark=*, mark size=2, mark=square]
table {%
-1.25 0.0565843621399137
-1 0.024176954732727
-0.75 0.00848765432098943
-0.5 0.00257201646141766
-0.25 0.00102880658445389
0 0.000514403292096578
0.25 0.000771604939494033
0.5 0.00128600823004736
0.75 0.00257201646061263
1 0.00668724279835199
1.25 0.0141460905349801
};
\addlegendentry{\textit{combined}-flow}

\end{axis}
\end{tikzpicture}
    \vspace{-0.5cm}
    \caption{EER as a function of the creak manipulation factor $\beta$ for the \textit{base}, the \textit{adapted} and \textit{adapted-2} models}
    \vspace{-0.5cm}
    \label{fig:eer}
\end{figure}

To assess the impact of \textit{creak} manipulation on speaker identity, we synthesized 4,000 utterances $\vect{\tilde{x}}_{n,\beta}$ from the test set at different \textit{creak} levels $\beta$, and extracted speaker embeddings using the same encoder as in the \gls{tts} system. Cosine similarities were computed for same-speaker (different utterances) and different-speaker pairs, while \gls{eer} served as the verification metric. All comparisons were made against speaker embeddings from unmanipulated utterances $\vect{x}_n$. \Cref{fig:eer} shows that manipulations with the \textit{base}-flow model have a much stronger impact on the \gls{eer} compared to the other systems. This suggests that stronger adjustments with this model lead to greater distortion of speaker characteristics, likely due to its pitch dependency. Some change in speaker identity is expected, since \textit{creak} is an inherent property of a speaker’s voice. However, manipulations with the \textit{adapted}-flow and \textit{combined}-flow models achieve a substantially better preservation of speaker identity across all manipulation levels, with virtually no difference between the two setups. Overall, speaker verification remains far more stable in models trained on the adapted datasets, even under strong \textit{creak} manipulations.

\section{Conclusion}
Training the normalizing flow for \textit{creak} manipulation on LibriTTS-R introduced a strong pitch bias, which degraded speaker verification after manipulation. Using data augmentation to remove the ``inter-speaker'' pitch–creak correlation improved verification performance, while the acoustic correlates still changed appropriately with the intended \textit{creak} level. Thus, \textit{creak} can be manipulated more effectively without compromising speaker identity.

\section*{Acknowledgement}
Funded by the Deutsche Forschungsgemeinschaft (DFG, German Research Foundation): TRR 318/1 2021- 438445824.

\newpage

\begingroup
\setstretch{0.89} %
\bibliographystyle{IEEEbib}
\bibliography{refs}
\endgroup

\end{document}